\documentclass[aps,superscriptaddress,twocolumn,twoside,floatfix,pra,nofootinbib,a4paper]{revtex4}
\usepackage{times}
\usepackage{epsfig}
\usepackage{amsfonts}
\usepackage{amsmath}
\usepackage{amssymb}
\usepackage{amsthm}
\usepackage{color}
\usepackage{enumerate}
\usepackage{multirow}
\usepackage[normalem]{ulem}
\newcommand{\stkout}[1]{\ifmmode\text{\sout{\ensuremath{#1}}}\else\sout{#1}\fi}
\usepackage{latexsym}
\usepackage{mathrsfs}
\usepackage{natbib}
\usepackage{verbatim}
\usepackage[T1]{fontenc}
\usepackage{float}
\usepackage{pifont}
\usepackage{mathtools}
\DeclarePairedDelimiter{\ceil}{\lceil}{\rceil}
\usepackage{lineno}

\usepackage{graphicx}
\usepackage{xcolor}

\usepackage[colorlinks=true,linkcolor=blue,citecolor=magenta,urlcolor=blue]{hyperref}

\definecolor{maroon}{cmyk}{0,0.87,0.68,0.32}

\newcommand{\ket}[1]{|#1\rangle}

\newcommand{\ketbra}[2]{|#1\rangle\langle#2|}

\newcommand{\comm}[1]{\textcolor{black}{#1}}

\begin{document}


\title{Demonstrating the power of quantum computers, certification of highly entangled measurements and scalable quantum nonlocality}

\author{Elisa B\"aumer}
\email{ebaeumer@itp.phys.ethz.ch} 
\affiliation{Institute for Theoretical Physics, ETH Zurich, 8093 Z\"urich, Switzerland}
\affiliation{IBM Quantum, IBM Research -- Zurich, 8803 R\"uschlikon, Switzerland}
\author{Nicolas Gisin}
\affiliation{D\'epartement de Physique Appliqu\'ee, Universit\'e de Gen\`eve, CH-1211 Gen\`eve, Switzerland}
\affiliation{Schaffhausen Institute of Technology -- SIT, Geneva, Switzerland}
\author{Armin Tavakoli}
\affiliation{D\'epartement de Physique Appliqu\'ee, Universit\'e de Gen\`eve, CH-1211 Gen\`eve, Switzerland}
\affiliation{Institute for Quantum Optics and Quantum Information -- IQOQI Vienna, Austrian Academy of Sciences, Boltzmanngasse 3, 1090 Vienna, Austria}

\begin{abstract}
Increasingly sophisticated quantum computers motivate the exploration of their abilities in certifying genuine quantum phenomena. Here, we demonstrate the power of state-of-the-art IBM quantum computers in correlation experiments inspired by quantum networks. Our experiments feature up to 12 qubits and require the  implementation of paradigmatic Bell-State Measurements for scalable entanglement-swapping. First, we demonstrate quantum \comm{correlations that defy classical models} in up to nine-qubit systems while only assuming that the quantum computer operates on qubits. Harvesting these quantum advantages, we are able to certify 82 basis elements as entangled in a 512-outcome measurement. Then, we relax the qubit assumption and consider quantum nonlocality in a scenario with multiple independent entangled states arranged in a star configuration. We report quantum violations of source-independent Bell inequalities for up to ten qubits. Our results demonstrate the  ability of quantum computers to outperform classical limitations and certify scalable entangled measurements. 
\end{abstract}

\maketitle


\section{Introduction}

Quantum computers have developed rapidly in recent years, with remarkable improvements in control, quality and scale.  While the presently available quantum computers have been used for realising many protocols and algorithms in quantum theory, it is interesting and important to consider the ability of such devices to realise predictions of quantum theory that cannot be explained by any conceivable classical model. A hallmark example of genuine quantum predictions is the violation of Bell inequalities, which has been demonstrated amongst others in a five-qubit transmon quantum computer \cite{Alsina2016} and a 14-qubit ion-trap quantum computer \cite{Lanyon2014}. 

The last decade has seen much attention directed at more sophisticated correlation experiments performed in quantum networks. These networks feature several parties that are connected through a given topology which may feature entangled states or quantum communication channels. Quantum networks are increasingly becoming practically viable \cite{Liao2018, Yin2020}, they have impactful potential applications \cite{Kimble2008, Wehner2018} and they raise new  conceptual questions.    The crucial conceptual feature, which sets quantum networks  apart from classical networks, is that initially independent entangled states distributed within the network can become globally entangled via the procedure of entanglement-swapping. Therefore, in contrast to e.g.~traditional Bell experiments, entangled measurements (i.e.~projections of several distinct qubits onto an entangled basis) are indispensable to  understanding and realising quantum correlations in networks. In particular, the paradigmatic Bell-State Measurement (BSM), known from quantum teleportation \cite{Teleport} and entanglement-swapping \cite{Swapping}, is at the heart of many schemes for quantum correlations in networks (see e.g.~\cite{Branciard2010, Branciard2012, Tavakoli2014, Tavakoli2016}). For the simplest network, quantum nonlocality has recently been experimentally demonstrated on optical platforms \cite{Saunders2017, Carvacho2017, Sun2019}. 

Here, we explore the ability of IBM quantum computers to realise quantum correlations that both defy classical models and certify entangled operations. 
\comm{These experiments can be seen as simulations of quantum networks as they realise the quantum correlations required for a quantum network on a single device.} 
\comm{While previous experiments with optical setups have demonstrated e.g.~nonlocality in the simplest network based on entangled measurements, superconducting circuits provide a platform where advanced quantum networks based on multi-qubit entangled measurement can be implemeneted. Due to free access to their superconducting quantum computers, IBM offers a convenient platform for such exploration \cite{IBM}. Currently there are in total 19 quantum devices consisting of up to 65 superconducting qubits on the IBM cloud, with around half of them being publicly available. Sophisticated quantum circuits can be built through the IBM Quantum Lab with Qiskit \cite{Qiskit}, which is an open-source software development kit for working with quantum computers at the level of pulses, circuits and application modules.} 

\begin{figure}
	\centering
	\includegraphics[width=0.9\columnwidth]{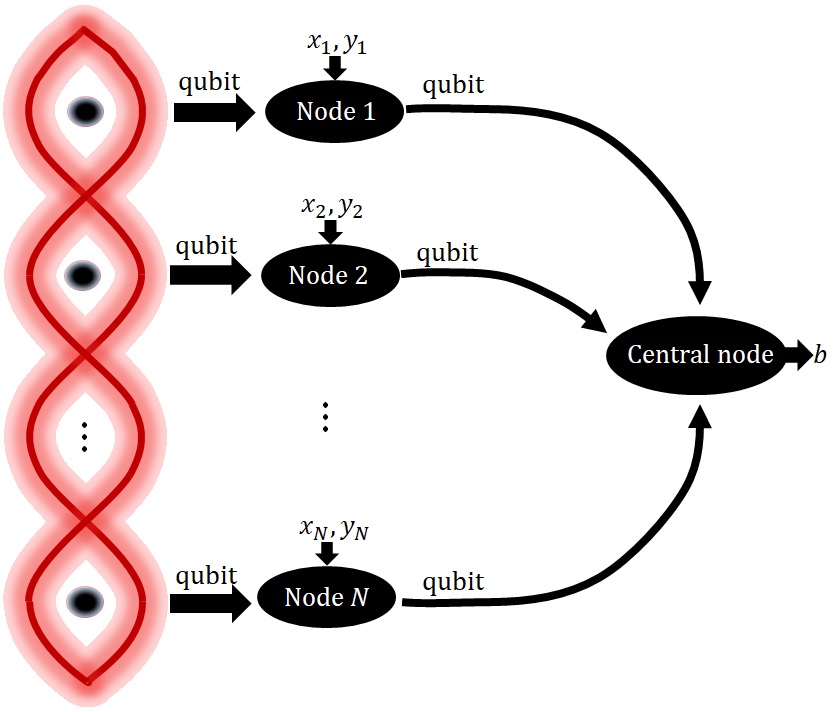}
	\caption{\textbf{Communication network.} Independent nodes share an $N$-qubit entangled state on which they perform qubit transformations. The shares are communicated to a central node which deterministically accesses information about the collective inputs of the $N$ other nodes by performing an $N$-qubit BSM. This gives rise to a quantum-over-classical advantage which enables a certification of the degree of entanglement present in the measurement of the central node.}\label{FigScenario2}
\end{figure}

In this work, we focus on two qualitatively different networks. Firstly, we consider a task in which entangled measurements are used to enhance \comm{ quantum correlations}  beyond the limitations of classical protocols \cite{FakeTriangle}. In this task, $N$ nodes share an $N$-qubit state and perform qubit transformations of their shares which they then relay to a final node that performs an $N$-qubit BSM (see Figure~\ref{FigScenario2}). The magnitude of the quantum-over-classical  advantage serves to certify the degree of entanglement present in the measurement under the sole assumption that  the quantum computer operates on qubits. We report quantum advantages for $N=2,\ldots,9$ but fail to observe an advantage for $N=10$. Importantly, for $N=9$, we can certify large-scale entanglement: our initially uncharacterised 512-outcome measurement has at least $82$ entangled basis elements. Secondly, we consider the  so-called star network in which a central node separately shares entanglement with $N$ initially independent nodes (see Figure~\ref{FigScenario1}). By implementing an $N$-qubit BSM in the central node, global entanglement can be established and quantum nonlocality can be demonstrated by violating a network Bell inequality \cite{Tavakoli2014}.  We report  quantum violations for $N=2,3,4,5$ while for $N=6$ we are unable to find a violation.  Finally, we go beyond BSMs  and consider correlation experiments based on the recently proposed quantum Elegant Joint Measurement \cite{EJM2019}. We realise this two-qubit entangled measurement and violate the bilocal Bell inequalities of Ref.~\cite{Tavakoli2020}. We also present a realisation of a quantum protocol in triangle-shaped configuration.

\section{Results}
\subsection{Certification of entangled measurements in communication network}
Consider the communication network illustrated in Figure~\ref{FigScenario2}. An $N$-qubit state is distributed between $N$ separate nodes. Each node receives an independent input corresponding to two bits $x_k,y_k\in\{0,1\}^2$, for $k=1,\ldots,N$, and implements a transformation of the incoming qubit. The transformed qubits are then communicated to a central node where they are collectively measured. The $N$-qubit measurement has $2^N$ possible outputs labelled by the bit-string $b\equiv b_1\ldots b_N\in\{0,1\}^N$. In this network, the state, transformations and measurement are uncharacterised up to an assumption of a qubit Hilbert space. The task is for the central node to recover knowledge about the input set $\{x_k,y_k\}_k$. Specifically, a successful information retrieval corresponds to  
\begin{equation}\label{win}
b_1=\bigoplus_{k=1}^N x_k \qquad \text{and}\qquad b_k=y_k\oplus y_1,
\end{equation}
for $k=2,\ldots,N$. Thus, only one of the $2^N$ possible outcomes of the measurement is considered successful.

Quantum theory offers a solution to the task \cite{FakeTriangle}.  Let the nodes share the $N$-qubit Greenberger-Horne-Zeilinger (GHZ) state $\ket{\text{GHZ}}=\frac{1}{\sqrt{2}}\left(\ket{0}^{\otimes N}+\ket{1}^{\otimes N}\right)$  and each perform local unitary operations corresponding to either $\openone$, $\sigma_X$, $\sigma_Z$ or $\sigma_Y$ (denoting the Pauli observables). The central node then performs an $N$-qubit BSM, i.e.~a projection onto a basis of GHZ-like states
\begin{equation}\label{GHZ}
\ket{M_b}=\sigma_Z^{b_1}\otimes \sigma_X^{b_2}\otimes \ldots\otimes \sigma_X^{b_N}\ket{\text{GHZ}},
\end{equation} 
whose outcome is guaranteed to satisfy the winning condition \eqref{win}. However, classical models that rely on the communication of binary messages can still partially perform the task. Denoting the average probability of satisfying all winning conditions \eqref{win} by $p_N^{\text{win}}$, the following limitations apply to classical and restricted quantum models \cite{FakeTriangle}: 
\begin{equation}\label{ineqs}
p_N^\text{win}\stackrel{\substack{\text{Classical}}}{\leq }\frac{1}{2} \stackrel{\substack{\text{\tiny At most m entangled}\\\text{\tiny measurement operators}}}{\leq}\frac{1}{2}\left(1+\frac{m}{2^N}\right) \stackrel{\text{Quantum}}{\leq} 1.
\end{equation}
\comm{Theoretically one can  reach $p_N^\text{win}=1 \ \forall N$ with the aforementioned quantum protocol. However,} every $p_N^\text{win}>1/2$ \comm{already} constitutes \comm{a proof of stronger-than-classical correlations}. \comm{This quantum correlation advantage is an advantage in communication complexity, where using a restricted amount of communication leads to a superior performance in terms of communication efficiency when using quantum instead of classical resources.} 

Furthermore, the second inequality in \eqref{ineqs} constitutes a bound respected by every possible quantum protocol in which the measurement in the central node has at most $m$ entangled basis elements. Therefore, the magnitude of the quantum advantage also determines a lower bound on the number of operators in the measurement that are certified as entangled, \comm{which is given by
\begin{align}
m = \ceil*{(2p_N^\text{win}-1)2^N}.
\end{align}
}

\begin{table}[t!]
	\begin{tabular}{|c|ccccccccc|}
		\hline
		N: \#qubits                                                                 & 2     & 3     & 4     & 5     & 6     & 7     & 8      & 9   & 10\\ \hline
		\begin{tabular}[c]{@{}c@{}}Measured $p_N^\text{win}$   (\%)\end{tabular}         & 93.9 & 89.2 & 85.0 & 80.4 & 73.5 & 67.5 & 63.7 & 58.0 &  $\tiny<\frac{1}{2}$ \\ \hline
		\begin{tabular}[c]{@{}c@{}}\#certified entangled \\ basis elements\end{tabular} & 4     & 7     & 12    & 20    & 30    & 45    & 70     & 82 & 0 \\ \hline
	\end{tabular}
	\caption{Results for $N$-qubit communication network experiments on the \emph{ibmq$\_$montreal} quantum computer. For every $N=2,\ldots,9$ we measure a better-than-classical success probability in the communication task while for $N=10$ we find no quantum advantage. As the number of qubits increases, the magnitude of the quantum advantage decreases. The number of certifiably entangled measurement operators peaks at $N=9$.}\label{TableFake}
\end{table}

We have implemented the quantum protocol for $N=2,\ldots,10$ with Qiskit \cite{Qiskit} on the \emph{ibmq$\_$montreal $27$-qubit} quantum computer \cite{IBM}. The results of the experiments are presented in Table~\ref{TableFake}. The standard deviation in $p_N^\text{win}$ is no larger than $10^{-3}$ for $N<10$. We find a quantum-over-classical advantage for $N=2,\ldots, 9$  but not for $N=10$. For the simplest case ($N=2$), we obtain a large quantum advantage and certify all four basis elements as entangled. As expected, the quantum advantage decreases as $N$ increases. Interestingly, however, the reduction is nearly linear (around five percentage points for each subsequent $N$) which attests to the scalability of the operations. Due to the slow decrease in the quantum advantage, we can report increasingly large entanglement certification: the most sizable entanglement is certified for $N=9$. In this case, our initially uncharacterised nine-qubit measurement has 512 possible outcomes. Via Eq.~\eqref{ineqs}, we can certify that at least $82$ measurement operators must be entangled. \comm{For $N=10$ we were not able to find any quantum advantage. The main reason is that a larger $N$ means a larger number of qubits in the protocol, which then requires larger number of CNOT gates and an increased circuit depth. This leads to an increased noise accumulation in the protocol. An additional reason is that on a $27$-qubit quantum computer some qubits and some connections perform better than others. While the provided average errors allow for a good estimate, it is still not straightforward which qubits to use in the experiment to optimise $p_{N}^\text{win}$. For $N<10$, we considered several different qubit architectures and repeated the experiments many times. However, the experiment for $N=10$ has over a million circuits and took approximately three days to perform. Therefore, we could not repeat it as many times in search of a large $p_{N}^\text{win}$.} As also noted in previous works \cite{Alsina2016}, we remark that the quantum computers typically are not stable as the results of our experiments vary significantly in time on otherwise identical implementations.

Furthermore, in the Supplementary Information we re-examine our results after applying measurement error-mitigation, which allows us to amplify the measured value of $p_N^\text{win}$.

\subsection{Star network nonlocality}

\begin{figure}
	\centering
	\includegraphics[width=0.9\columnwidth]{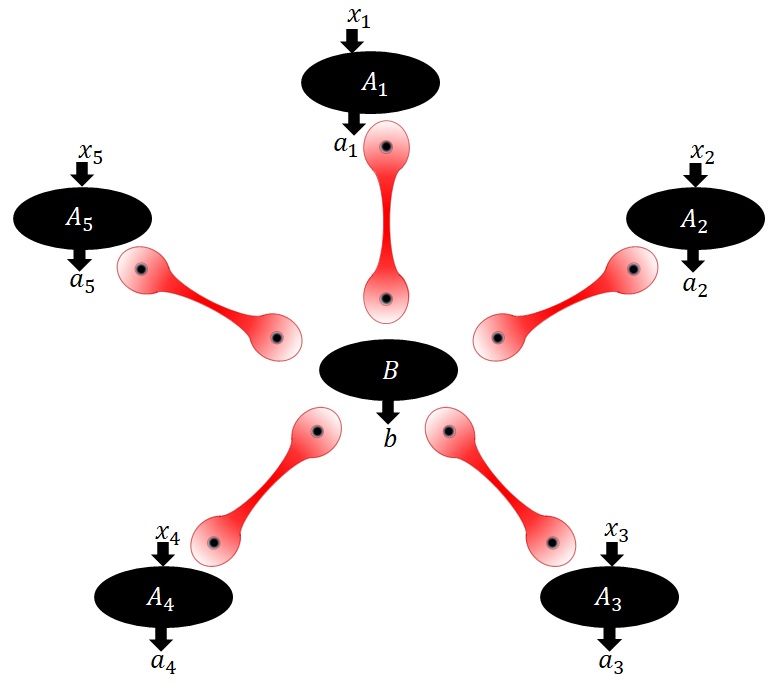}					
	\caption{\textbf{Star network.} A central node independently shares five pairs of entangled qubits with separate nodes. By performing a BSM on  five qubits, the central node renders the five initially independent nodes in a globally entangled state. With suitable local measurements, the correlations in the network become nonlocal.}\label{FigScenario1}
\end{figure}

We proceed to consider a network of the type illustrated in Figure~\ref{FigScenario1}. A central node, $B$, is connected to $N$ other nodes, $A_1,\ldots, A_N$, through independent sources emitting pairs of particles. The branch-nodes each have independent binary inputs $\bar{x}\equiv x_1,\ldots,x_N\in \{0,1\}$ and produce binary outputs $\bar{a}\equiv a_1,\ldots,a_N\in\{0,1\}$ while the central node has a fixed setting and produces an output $b\equiv b_1\ldots b_N\in\{0,1\}^N$ that can take $2^N$ different values. The probability distribution in the network is written $p(\bar{a},b|\bar{x})$.  It is said to admit a local model, that respects the independence of the $N$ sources, if it can be written in the form 
\begin{equation}
p(\bar{a},b|\bar{x})=\int d\bar{\lambda} \left(\prod_{i=1}^{N}q_i(\lambda_i)p(a_i|x_i,\lambda_i)\right)p(b|\bar{\lambda}),
\end{equation}
where $\lambda_i$ is the local variable associated to the $i$'th source, $q_i$ is its probability density and $\bar{\lambda}=(\lambda_1,\ldots,\lambda_N)$. Ref.~\cite{Tavakoli2014} introduced Bell inequalities respected by all source-independent local models:
\begin{equation}\label{starineq}
\mathcal{S}_N\equiv \frac{1}{2^{N-2}}\sum_{j=1}^{2^{N-1}} |I_j|^{1/N}\leq 1,
\end{equation} 
where $I_1,\ldots, I_{2^{N-1}}$ are suitable linear combinations of $p(\bar{a},b|\bar{x})$. See Supplementary Information for further details on the inequalities. A quantum violation of \eqref{starineq} is possible if each source distributes a maximally entangled two-qubit state $\ket{\psi}=\frac{\ket{00}+\ket{11}}{\sqrt{2}}$, the branch-nodes $A_1,\ldots, A_N$ measure the observables $\frac{\sigma_X+\sigma_Y}{\sqrt{2}}$ and $\frac{\sigma_X-\sigma_Y}{\sqrt{2}}$ and the central node performs the $N$-qubit BSM \eqref{GHZ}.This leads to the quantum violation $\mathcal{S}_N=\sqrt{2}$ for every $N$. \comm{Note, that $N$ corresponds to the number of branches, and it implies that the quantum setup is implemented on $2N$ qubits.}

\begin{table}[]
	\begin{tabular}{|c|ccccc|}
		\hline
		N: \#branches in network             & 2     & 3     & 4     & 5   &  6  \\ \hline
		\begin{tabular}[c]{@{}c@{}}Measured $\mathcal{S}_N$\end{tabular}         & 1.165 & 1.124 & 1.086 & 1.062 & 0.983  \\ \hline
		\begin{tabular}[c]{@{}c@{}} KL-divergence  \end{tabular} & 4.1 e-6  &  1.3 e-3    & 2.8 e-4   &  2.5 e-3 &  1.7 e-3 \\ \hline
	\end{tabular}
	\caption{Results for $N$-branch star network nonlocality experiments on the \emph{ibmq$\_$almaden} quantum computer. The source-independent local bound is violated for $N=2,3,4,5$ but not for $N=6$. With increasing number of qubits ($2N$ qubits in the network), the magnitude of the violation decreases. The Kullback-Leibler divergence, employed to estimate the accuracy of the source independence assumption, remains small even at large $N$.}\label{TabStar}
\end{table}

We have implemented this quantum protocol with Qiskit \cite{Qiskit} on the \emph{ibmq$\_$almaden $20$-qubit} quantum computer \cite{IBM} for star networks with $N=2,3,4,5,6$. The results of the experiments are presented in Table~\ref{TabStar}. The standard deviations associated to the statistical fluctuations for $\mathcal{S}_N$ for $N=2,\ldots,6$ are in all cases smaller than $2\times 10^{-3}$. We find a quantum violation of the source-independent local bound for $N=2,3,4,5$ but fail to observe a violation for $N=6$. \comm{With an increasing number of qubits, the size of the joint measurement increases and therefore also the noise. When considering the nearly linear decrease of the measured $\mathcal{S}_N$ values for  $N=2,3,4,5$, it is not unexpected that for $N=6$ the value falls below threshold $\mathcal{S}_N=1$.} For the simplest case ($N=2$), the quantum violation is already far from the theoretical maximum of $\sqrt{2}$, which attests to the demanding nature of falsifying source-independent local models. However, the relatively small decrease in the violation magnitude for each subsequent $N$ attests to the scalability of the experiment.

Moreover, in our proof-of-principle demonstration, the sources are not perfectly independent, e.g.~due to cross talk \cite{Gambetta2012, Takita2017, Mundada2019}. We have estimated the degree of source-independence in our experiments by evaluating the worst-case Kullback-Leibler (KL) divergence (i.e.~the relative entropy,  maximised over all settings $\bar{x}$) between the marginal distribution $p(\bar{a}|\bar{x})$ and would-be marginal distribution had the sources been perfectly independent; $\prod_{i=1}^N p(a_i|\bar{x})$. In Table~\ref{TabStar}, we see that the worst-case KL-divergence is nearly vanishing for $N=2$ and remains low also for larger $N$. 

Finally, in the Supplementary Information, we re-examine the results of our experiments after applying error-mitigating post-processing to the measured probabilities.

\subsection{Experiments based on the Elegant Joint Measurement}
While our main focus has been on Bell-State Measurements, there has recently been a proposal of another natural, yet qualitatively different, entanglement-swapping measurement that has a high degree of symmetry. This so-called Elegant Joint Measurement (EJM) \cite{EJM2019}, which projects two qubits onto a basis of partially (but equally) entangled states with a tetrahedral symmetry, has been placed at the heart of several protocols for quantum networks. 
For instance, Ref.~\cite{Tavakoli2020} considered the simplest star network ($N=2$) and proposed a source-independent Bell inequality tailored to be violated with the EJM performed in the central node. This test of quantum correlations is conceptually different and practically more challenging than the previously considered scenario based on the BSM: the quantum correlations are significantly more fragile to imperfections and the circuit implementation of the EJM requires more entangling gates. \comm{In Ref.~\cite{Tavakoli2020}, it was shown that the EJM can be implemented by using Hadamard gates $H=\frac{1}{\sqrt{2}}\sum_{i,j=0,1}(-1)^{ij}\ketbra{i}{j}$, phase gates $S:=R_z(\pi/2)=\ketbra{0}{0}+i\ketbra{1}{1}$, controlled phase gates $CR_z(\pi/2)=\ketbra{0}{0}\otimes \openone+\ketbra{1}{1}\otimes S$ and controlled not gates $CNOT=\ketbra{0}{0}\otimes \openone+\ketbra{1}{1}\otimes \sigma_X$ in the following configuration:
\begin{equation}
\ket{\text{EJM}_{k}}=CNOT(H\otimes \openone)CR_z(\pi/2)(S \otimes S)(H\otimes H)\ket{k_1k_2},
\end{equation}
for $k\equiv k_1k_2\in\{0,1\}^2$. We implemented the EJM in the context of two experiments: a test of the bilocal Bell inequalities of Ref.~\cite{Tavakoli2020} and a test of the quantum triangle network discussed in Ref.~\cite{EJM2019}.}

\subsubsection{The bilocal scenario}
\comm{
The bilocal scenario is the simplest star network (corresponding to $N=2$), i.e., it features three nodes in a line configuration where the first and second as well as the second and third are connected. The two branch-nodes each receive inputs $x,z \in \{1,2,3\}$, respectively, which determine the bases in which they perform their measurements, while the central node performs always the same measurement. They obtain outputs $a,c \in \{1,-1\}$ and $b \in \{1,2,3,4\}$, respectively. From the resulting conditional probability distribution $p(a,b,c|x,z)$, one can then determine the conditional single-party correlators $E_b^A(x) := \frac{1}{3} \sum_{a,c,z} a\ p(a,c|b,x,z)$ and  $E_b^C(z) := \frac{1}{3} \sum_{a,c,x} c\ p(a,c|b,x,z)$ as well as the conditional two-party correlator $E_b^{AC}(x,z) := \sum_{a,c} ac\ p(a,c|b,x,z)$. Denote the four vertices of a tetrahedron by the coordinates $\vec{m}_1 = (+1, +1, +1)$, $\vec{m}_2 = (+1, -1, -1)$, $\vec{m}_3 = (-1, +1, -1)$, $\vec{m}_4 = (-1, -1, +1)$ and define $m_b^k$ as the $k$th element of $\vec{m}_b$. Ref.~\cite{Tavakoli2020} introduced the bilocal Bell inequality
\begin{multline}
\mathcal{B} := \sum_{x,b} \sqrt{p(b)(1-m_b^x E_b^A(x))} +  \sum_{z,b} \sqrt{p(b)(1+m_b^z E_b^C(z))}  \\
+ \sum_{x \neq z,b} \sqrt{p(b)(1-m_b^x m_b^z E_b^{AC}(x,z))}\\
\leq  12 \sqrt{3} + 2 \sqrt{15} \approx 28.531.
\end{multline}
In the quantum setup the two sources each distribute singlet states $\ket{\psi^-}=\frac{\ket{01}-\ket{10}}{\sqrt{2}}$, the central node performs the EJM and the two branch-nodes perform the measurements $\sigma_X, \sigma_Y, \sigma_Z$. Theoretically this gives the violation $\mathcal{B}=12\sqrt{6}\approx 29.39$. Running the experiment on the \emph{ibmq$\_$manhattan $65$-qubit} quantum computer, the resulting probability distribution yielded the value $\mathcal{B} = 28.648 \pm 0.008$. This constitutes a small but statistically significant violation of the source-independent local bound.} 

\begin{figure}
	\centering
	\includegraphics[width=0.9\columnwidth]{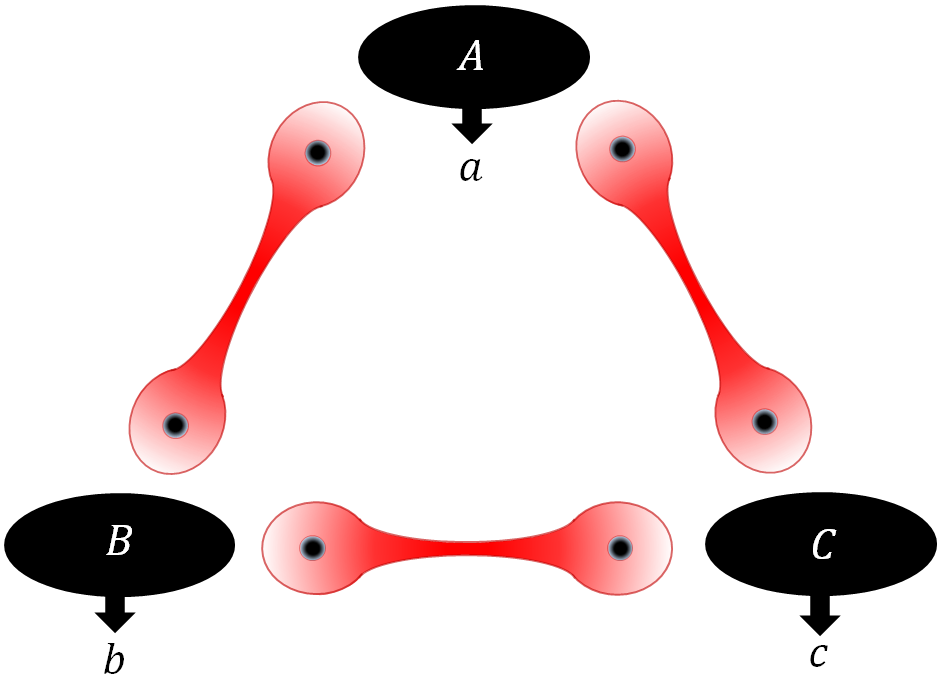}
	\caption{\textbf{Triangle network.} Three nodes pairwise share pairs of entangled qubits and each perform the Elegant Joint Measurement on their respective qubit pair.}\label{FigTriangle}
\end{figure}

\subsubsection{Quantum triangle network}
\begin{center}
\begin{figure*}[t!]
\centering
\includegraphics[width=2\columnwidth]{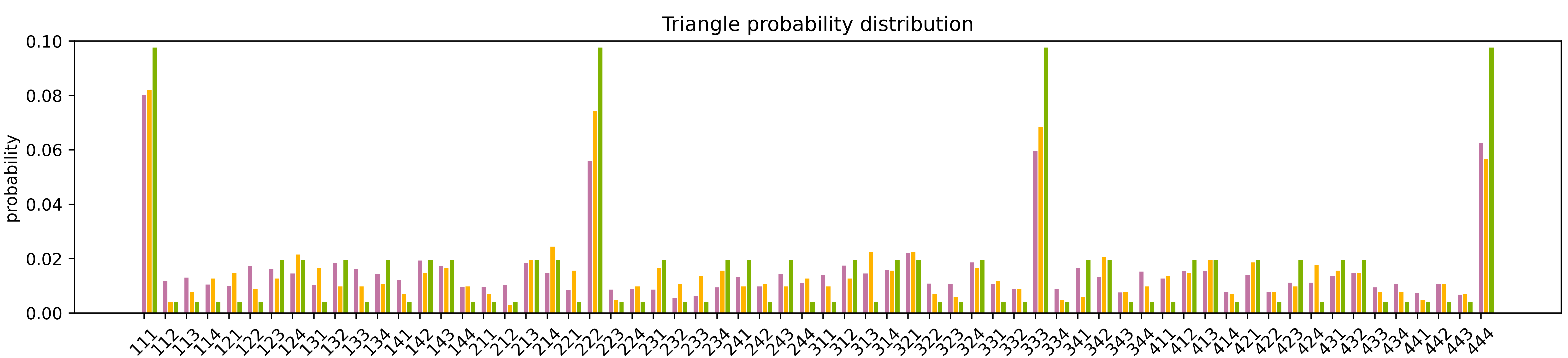}					
\caption{\textbf{Triangle network results.} \comm{Histogram of the probability distribution of the quantum triangle network. The green distribution corresponds to the theoretical prediction, while the purple and orange correspond to the outcomes of experiments conducted on the \textit{ibmq\_montreal} device and the \textit{ibmq\_johannesburg} device, respectively.}}
\label{FigHistTriangle}
 \end{figure*}
 \end{center}

Furthermore, it is conjectured that the EJM can reveal quantum correlations also in a triangle-shaped network  \cite{EJM2019},
\comm{i.e. a network featuring three nodes that are pairwise connected (see Figure~\ref{FigTriangle}). Each node performs a single measurement and obtains an output $a,b,c\in\{1,2,3,4\}$ respectively. The resulting probability distribution has a local model that respects the independence of the three sources if it can be written as
\begin{equation}
p(a,b,c)=\int d\alpha d\beta d\gamma p_A(a|\beta,\gamma)p_B(b|\alpha,\gamma)p_C(c|\alpha,\beta),
\end{equation}
where $(\alpha,\beta,\gamma)$ are local variables. While it is known that there exist quantum correlations that do not admit the above form \cite{Salman2019}, no noise-robust examples are presently known. It is, however, conjectured that noise-robust quantum correlations are obtained if all three sources emit a singlet state $\ket{\psi^-}=\frac{\ket{01}-\ket{10}}{\sqrt{2}}$ and all three nodes perform the EJM on the two independent qubits at their disposal. 
The corresponding probability distribution in the network is fully described by three cases: 
\begin{enumerate}[i)]
\item when all outcomes are equal we have $p(r,r,r)=\frac{25}{256}$  for $r=1,2,3,4$, 
\item when precisely two outcomes are equal we have $p(r,r,s)=\frac{1}{256}$ for all $r\neq s$ (including permutations of the labels),\item when all outcomes are different we have $p(r,s,t)=\frac{5}{256}$ for all $r\neq s \neq t \neq r$.
\end{enumerate}}
We have realised this quantum protocol \comm{on the \emph{ibmq$\_$montreal $27$-qubit} quantum computer and the \emph{ibmq$\_$johannesburg $20$-qubit} quantum computer. The two experimentally obtained probability distributions as well as the theoretical one are shown in Figure \ref{FigHistTriangle}. While the four peaks for $a=b=c$ are quite distinguishable, the noise prevents us from always clearly identifying the cases of two and three different outcomes. The coupling map of the \emph{ibmq$\_$johannesburg} device was better suited as we could perform the experiment on a ring of six qubits, while on the \emph{ibmq$\_$montreal} device we had to choose six qubits on a ring of twelve qubits (see SM). Nevertheless, as the error rates on the \emph{ibmq$\_$montreal} device are smaller, the KL-divergences between the experimental and theoretical distributions are very similar with $0.272$ for the resulting distribution of the \emph{ibmq$\_$johannesburg} device and $0.277$ for the one of the \emph{ibmq$\_$montreal} device.}

However, there presently exists no criterion for determining whether our measured correlations can be simulated in a source-independent local model. Determining whether our measured correlations elude all source-independent local models remains an open problem.

\section{Discussion}
We have reported demonstrations of quantum predictions that defy general classical models in scalable experiments featuring up to ten qubits in which the central component is the implementation of sophisticated entangled measurements of many qubits. Our experiments mirror the conditions encountered in quantum networks, but are rightfully viewed as simulations of networks since all physical qubits are confined to a single IBM quantum processor.  Our results demonstrate the power and scalability of these state-of-the-art devices, most notably allowing for the certification of a nine-qubit measurement with $82$ entangled basis elements. They also offer an avenue for asserting the quality of a quantum computer based on whether (and to what extent) it can generate quantum correlations that elude classical models. Finally, our results also indicate the prospects for real-life local area quantum networks based on transmon quantum computers in which physical distances can be mediated by microwave photons. \comm{While this \textit{quantum transduction} might still be a few years ahead, there have already been significant steps towards developing such a quantum-coherent, bidirectional transducer between microwave and optical frequencies \cite{Roch2014, Narla2016, Ibarc2018, Kurpiers2018, Krinner2019, Schneider2019, Mirhosseini2020}.} This would be an interesting endeavour for future research.

\section{Methods}
\subsection{Experimental Optimization}
All of the devices we used operate on superconducting transmon qubits located in a fridge with a temperature of roughly 15 mK. \comm{In order to find the most suitable devices, including the choice of which qubits to use, we have considered the connectivity maps to minimise the number of overhead CNOT gates, together with the performance of the individual qubits and connecting gates, such as the T1 and T2 times, the measurement errors and the CNOT gate fidelities. The quantum volume \cite{QVolume} is also a good indicator for the overall performance of the different devices. However, when only a small subset of qubits of a device is used, the fact that there is quite some variation between different qubits on the same device, makes the optimal choice not straightforward. The final optimisation was in our case performed empirically, i.e. by testing the most promising configurations and comparing them. We do not know of any rigorous and general method to find the best device, and suspect that this would be challenging since we observe that the performance of different devices varies substantially depending on the choice of circuits to be implemented.}
 
\subsection{Communication Network}
For the communication network, we considered several devices before choosing the \emph{ibmq$\_$montreal $27$-qubit} quantum computer due to its comparatively high performance. \comm{The relatively small two-qubit gate errors, large coherence times and small readout errors on up to ten connected qubits resulted in a larger success probability and therefore allowed us to certify more entangled operators.} Each experiment features $4^N$ circuits which are realised using a different set of local unitaries (exemplified in SM). The GHZ-state and the BSM are realised using a Hadamard gate followed by several CNOT gates acting pairwise between the $N$ qubits. Since different qubits on the device are subject to different gate errors, relaxation times and dephasing times, we have strived to choose our $N$ working qubits favourably. In order to enable the practical implementation of the exponentially growing number of circuits, we have exploited that the measured winning probability is constant in $N$. This allows us to exponentially reduce the number of shots per circuit while maintaining a low standard deviation for $p_N^\text{win}$ (see SM). Therefore, each circuit was implemented in 24576 shots ($N=2$), 8192 shots ($N=3,4$), 1024 shots ($N=5$), 128 shots ($N=6$), 32 shots ($N=7,8$), 16 shots ($N=9$) and 8 shots ($N=10$), leading to a standard deviation (statistical error) no larger than $10^{-3}$ in the estimate of $p_N^\text{win}$ for $N<10$ (see Supplementary Information for details).

\subsection{Star Network}
Similarly, for the star network, we also considered several IBM devices but focused on the \emph{ibmq$\_$almaden $20$-qubit} quantum computer due to its favourable qubit architecture, the good results and the fact that it supports the experiment for  $N=2,3,4,5,6$. \comm{Specifically, the connectivity allows for the six qubits that form the central node to be each connected to another branch node, such that no swap gates are needed to address the corresponding two-qubit gates.} The experiments require $2^N$ different circuits, each corresponding to a set of local measurements (exemplified in SM). In order to keep the standard deviation low, we performed an increasing number of shots per circuit as we increased $N$. For each circuit we have implemented approximately $1.2\times 10^5$ shots ($N=2,3$), $2\times 10^5$ shots ($N=4,5$) and $4.9\times 10^6$ shots ($N=6$), leading to a standard deviation no larger than $2\times 10^{-3}$ in the estimate of $\mathcal{S}_N$ for $N=2,\ldots,6$ (see Supplementary Information for details). 

\subsection{Bilocal Scenario}
\comm{For the bilocal scenario, we needed four qubits connected in a line that were notably robust. We run the experiment on all promising devices, but were only able to demonstrate a violation of the inequality of Ref.~\cite{Tavakoli2020} using the \emph{ibmq$\_$manhattan $65$-qubit} quantum computer. We ran $3.3 \times 10^5$ shots per setup.}

\subsection{Triangle Scenario}
\comm{Since there presently does not exist a noise-robust Bell-type inequality for the quantum correlations targeted in our experiment in the triangle scenario, it is not straightforward to determine how faithfully our result matches the theoretical quantum predictions. We considered the KL-divergence between the experimental and theoretical distribution and obtained the best results using the \emph{ibmq$\_$montreal $27$-qubit} quantum computer and the \emph{ibmq$\_$johannesburg $20$-qubit} quantum computer, which can be explained by their small readout and two-qubit gate errors and their suitable connectivity map, respectively.}

\section{Data Availability}
The data that support the findings of this study are available from the corresponding author upon reasonable request.

\section{Acknowledgements}
This work was supported by the Swiss National Science Foundation (Starting grant DIAQ, NCCR-QSIT, NCCR-SwissMAP, Early Mobility Fellowship P2GEP2 194800, as well as project No.\ 200021\_188541) and by the Air Force Office of Scientific Research via grant FA9550-19-1-0202. We acknowledge use of the IBM Quantum services \cite{IBM}.

\section{Competing Interests}
The authors declare no competing interests.

\section{Author Contributions}
A.T. developed the main ideas behind the project. E.B. ran and evaluated the experiments. A.T. wrote the main part of the manuscript. All authors discussed the results and contributed to the manuscript.

\appendix
\onecolumngrid

\section{Circuits and layout}\label{AppCircuit}
\begin{figure}[htb]
	\centering
	\includegraphics[width=0.6\columnwidth]{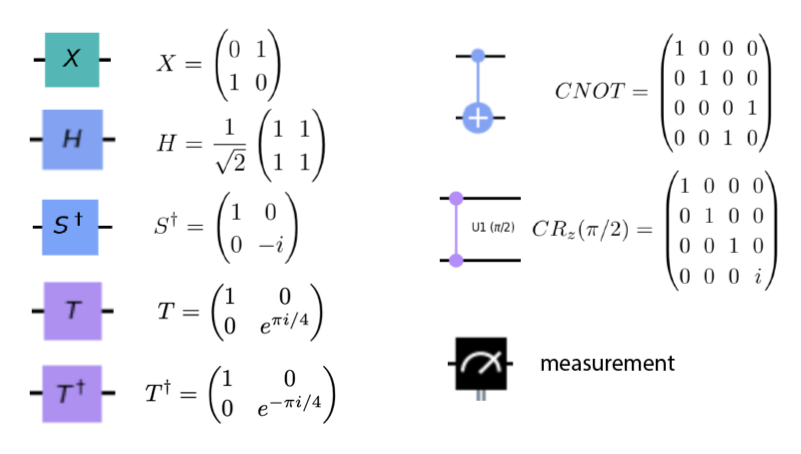}					
	\caption{\textbf{Gates.} Legend providing the matrix description of the different gates used in our experiments.}\label{FigGates}
\end{figure}

Figure \ref{FigGates} provides the mathematical description of all gates that appear in the circuit diagrams (below) describing our experiments.

\subsection{Communication network experiment} \label{AppCommCirc1}
\begin{figure}[H]
	\centering
	\includegraphics[width=0.65\columnwidth]{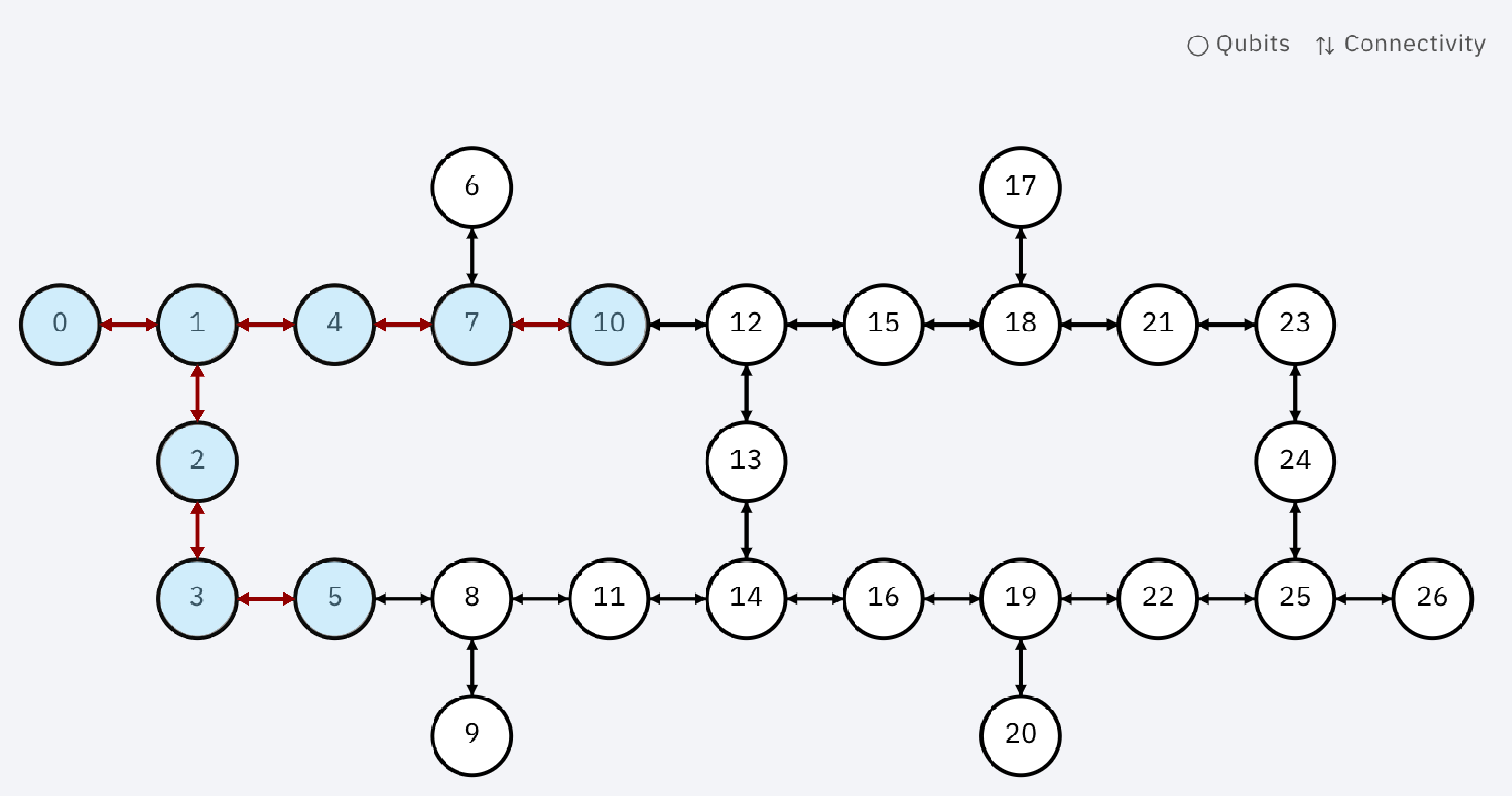}					
	\caption{\textbf{Connectivity map of the device \textit{ibmq\_montreal}.} An 8-qubit GHZ-state is created on the blue nodes, followed by local unitaries and a BSM. }\label{FigConnMontreal}
\end{figure}
\begin{figure}[h!]
	\centering
	\includegraphics[width=0.8\columnwidth]{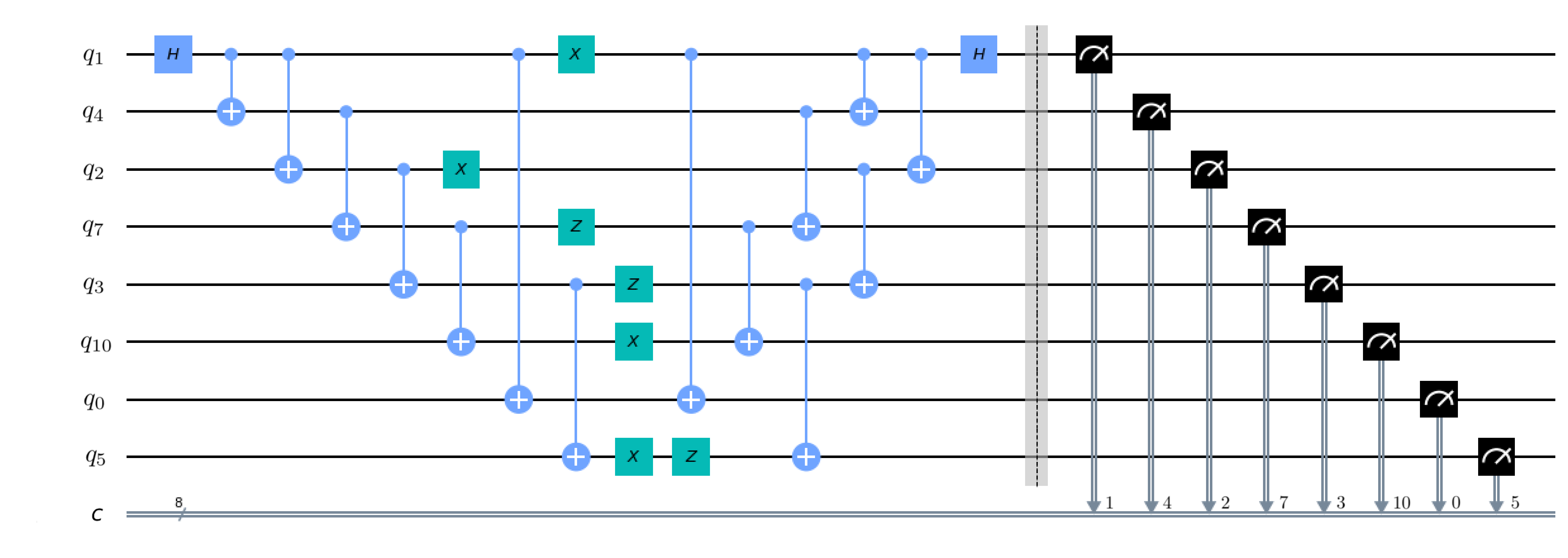}					
	\caption{\textbf{Circuit implementing the communication network experiment for $N=8$ qubits with the setup $x=(00011001)$, $y=(10100101)$.} \comm{The single lines labeled by $q_k$ correspond to the respective qubits $k$ that were used in the experiment, while the double line labeled by $c$ corresponds to the classical outputs that the measurements yield.} The Hadamard gate and the CNOT gates constitute the creation of a GHZ-state in the beginning and the BSM in the end, respectively, while the X and Z gates are the operations corresponding to the bits $\{x_k\}_{k=1}^N$ and $\{y_k\}_{k=1}^N$, respectively.}\label{FigCircMontreal}
\end{figure}

In Figure \ref{FigConnMontreal} the connectivity map of the \textit{ibmq\_montreal} device is displayed. This device was chosen for the communication network experiment because of its comparatively small readout and two-qubit gate errors. The particular colouring corresponds to the case of a communication network involving eight qubits. One exemplifying circuit, corresponding to a particular choice of settings $\{x_k,y_k\}_{k=1}^{N}$, is shown in Figure \ref{FigCircMontreal}.

\subsection{Star network experiment} \label{AppStarCirc}

\begin{figure}[htb]
	\centering
	\includegraphics[width=0.4\columnwidth]{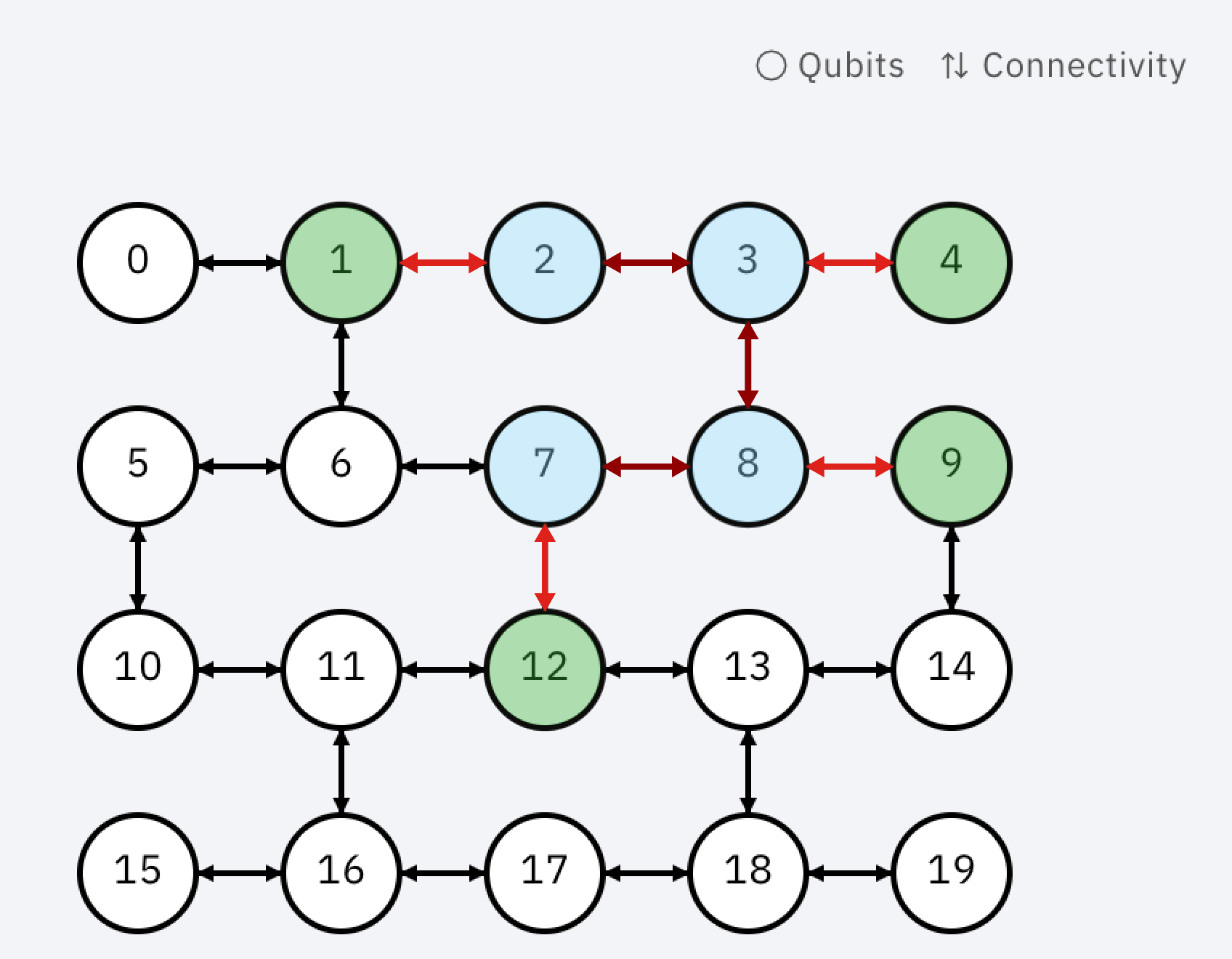}					
	\caption{\textbf{Connectivity map of the device \textit{ibmq\_almaden}.} The red arrows indicate the four separate maximally entangled qubit pairs. The blue qubits are held by the central node while the green ones represent the independent branch-nodes. This colouring corresponds to $N=4$.}\label{FigConnAlmaden}
\end{figure}

\begin{figure}[htb]
	\centering
	\includegraphics[width=0.5\columnwidth]{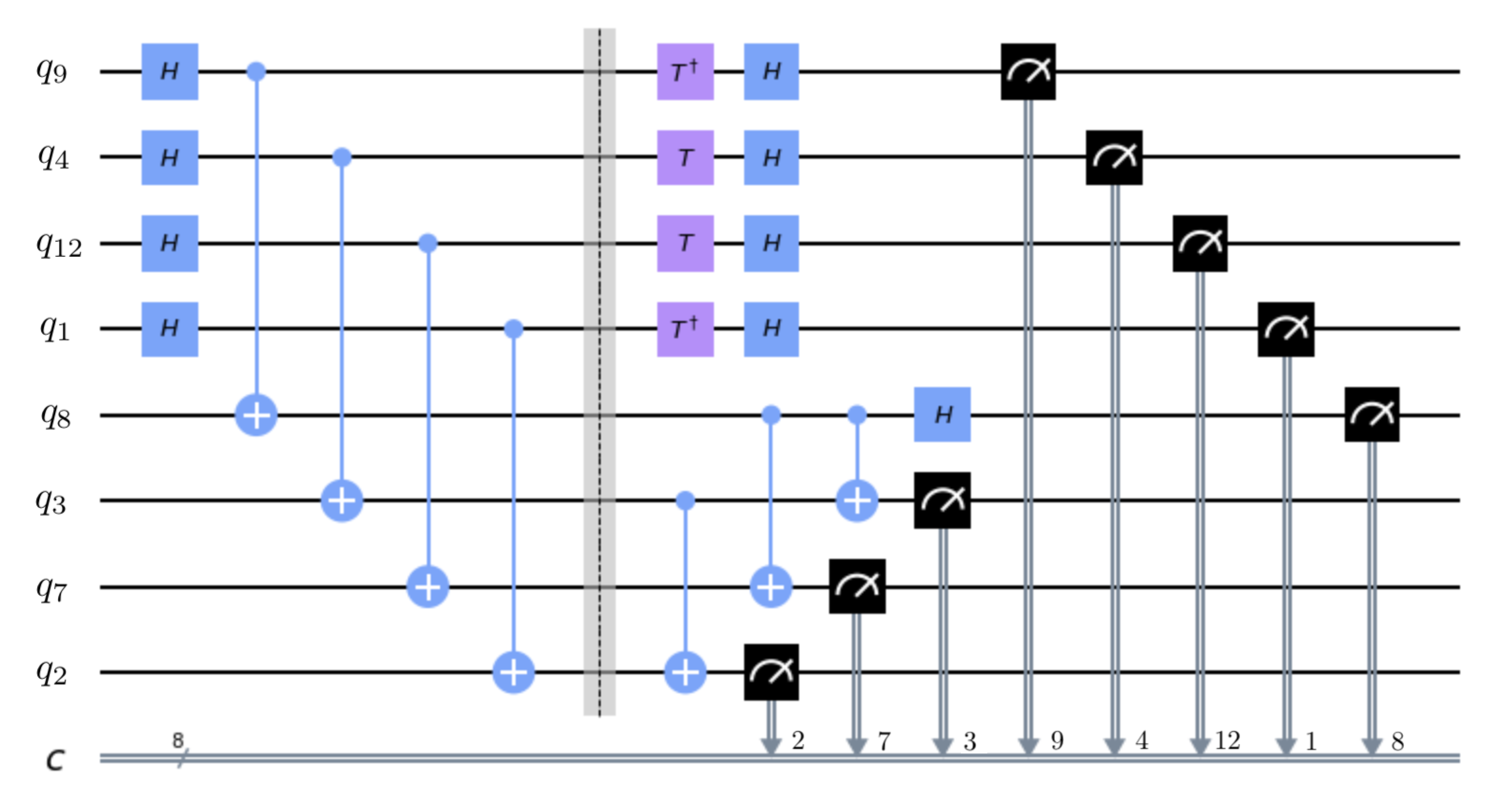}					
	\caption{\textbf{Circuit implementing the star network experiment for $N=4$ with the inputs $x_1=x_4 = 0$ and $x_2=x_3=1$.} We start by creating the maximally entangled state $\ket{\psi}=\frac{\ket{00}+\ket{11}}{\sqrt{2}}$ on each of the qubit pairs $(q_9, q_8)$, $(q_4, q_3)$, $(q_{12}, q_7)$, $(q_1, q_2)$ by applying a Hadamard gate followed by a CNOT gate. According to $x$ we then apply either a $T$-gate or a $T^\dag$-gate to each of the branch qubits $\{q_9, q_4, q_{12}, q_1\}$ to measure in the desired basis, while on the central node qubits $\{q_8, q_3, q_7, q_2\}$ we perform a BSM.}\label{FigCircAlmaden}
\end{figure}

In Figure~\ref{FigConnAlmaden} the connectivity map of the \textit{ibmq\_almaden} device is displayed. This device was chosen for the star network experiment because of its good connectivity which allowed each of the qubits that belonged to the ``central node'' to be connected to an additional ``branch-qubit''. The particular colouring displayed in the figure corresponds to the case of four sources in the network. One exemplifying circuit, corresponding to a particular choice of settings $\bar{x}$, is shown in Figure \ref{FigCircAlmaden}.

\subsection{Bilocality experiment} \label{AppBiloc}
\begin{figure}[H]
	\centering
	\includegraphics[width=0.65\columnwidth]{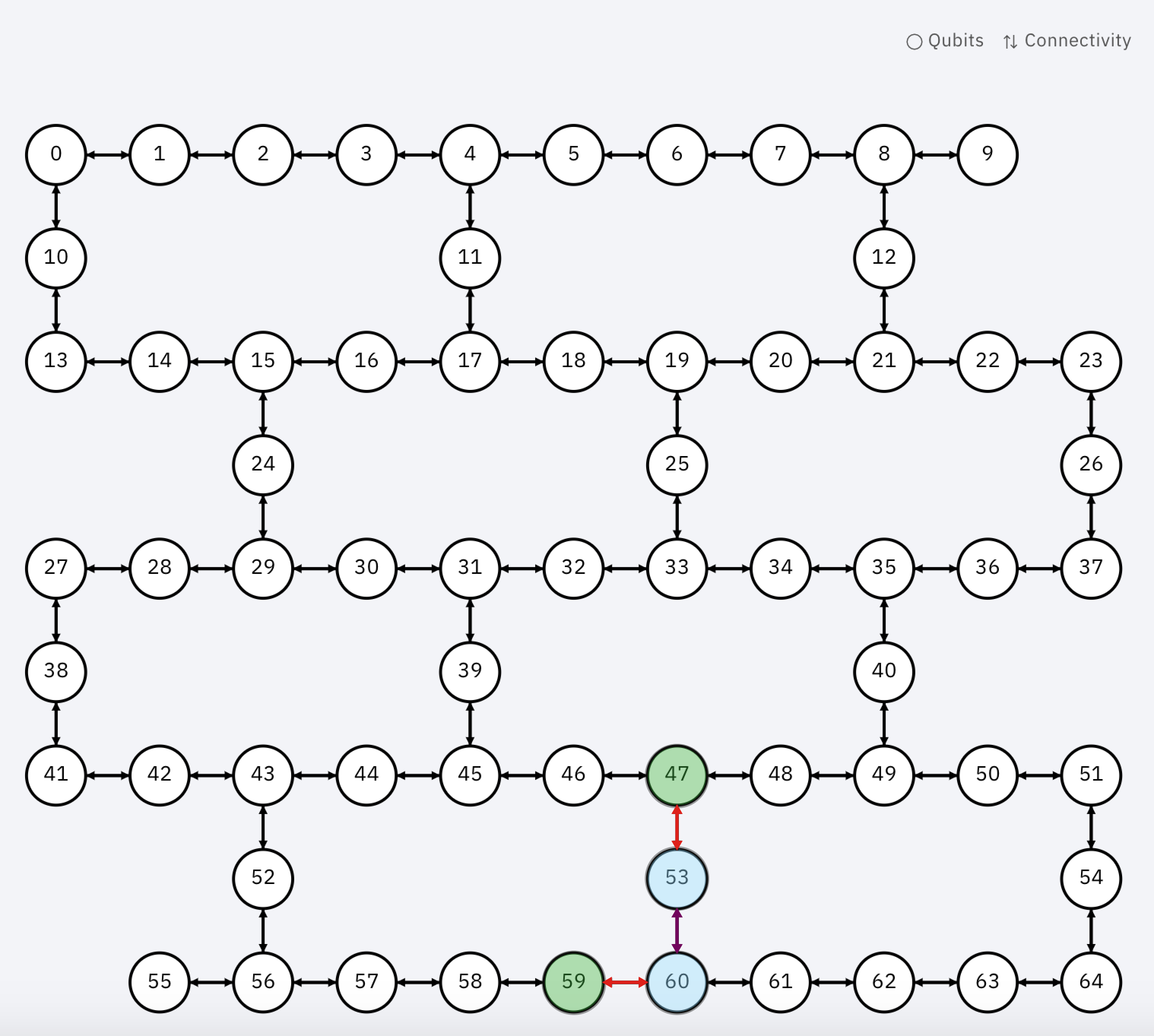}					
	\caption{\textbf{Connectivity map of the device \textit{ibmq\_manhattan}.} The red arrows indicate the two separate maximally entangled qubit pairs. The blue qubits are held by the central node while the green ones represent the two independent branch-nodes. }\label{FigConnManhattan}
\end{figure}

\begin{figure}[htb]
	\centering
	\includegraphics[width=0.6\columnwidth]{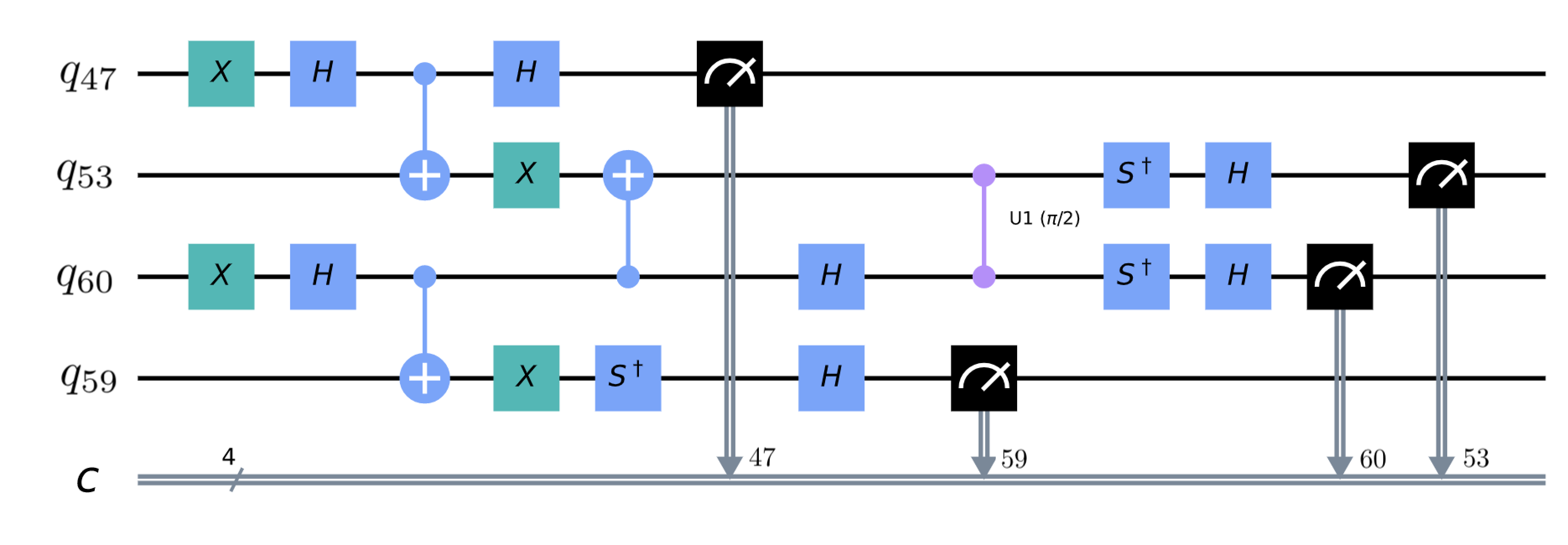}					
	\caption{\textbf{Circuit implementing the bilocality experiment on four qubits with the setup $x=1$, $z=2$.} We start by creating the maximally entangled singlet state $\ket{\psi}=\frac{\ket{01}-\ket{10}}{\sqrt{2}}$ on both qubit pairs $(q_{47}, q_{53})$ and $(q_{60}, q_{59})$ by applying an $X$-gate and a Hadamard gate followed by a CNOT gate and another $X$-gate on the second qubit. According to $x$ and $z$, we then apply an $H$-gate on $q_{47}$ and both, an $S^\dag$-gate and an $H$-gate on $q_{59}$ to measure in the desired bases, while on the central node qubits $\{q_{53}, q_{60}\}$ we conduct an EJM. Note, that the measurements are all conducted at the same time, even though this circuit might suggest otherwise.}\label{FigCircManhattan}
\end{figure}

In Figure \ref{FigConnManhattan} the connectivity map of the \textit{ibmq\_manhattan} device is displayed. This new device was chosen for the bilocality experiment because of its small readout and two-qubit gate errors. The particular colouring corresponds to the four qubits that were used in this experiment. One exemplifying circuit, corresponding to a particular choice of settings $\{x,z\}$, is shown in Figure \ref{FigCircManhattan}.

\comm{
\subsection{Triangle network experiment} \label{AppTriangle}
\begin{figure}[htb]
	\centering
	\includegraphics[width=0.8\columnwidth]{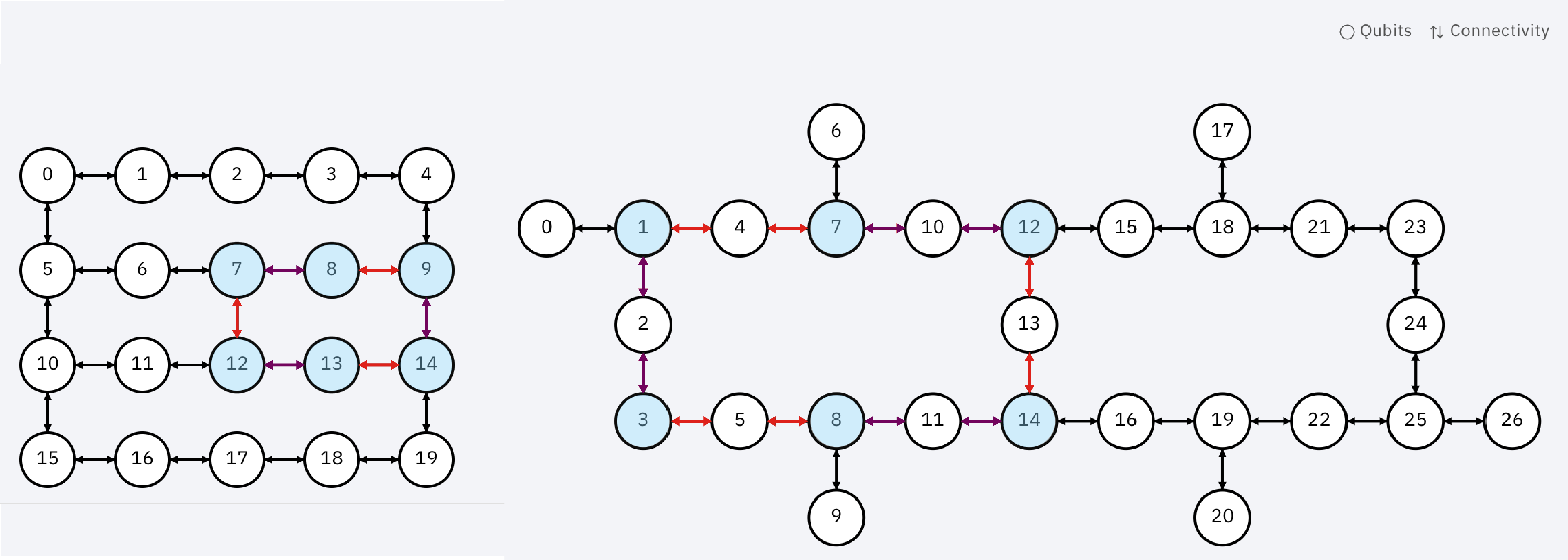}					
	\caption{\textbf{Connectivity maps of the devices \textit{ibmq\_johannesburg} and \textit{ibmq\_montreal}, respectively.} While the red connections link the qubits that share a maximally entangled state, the purple connections link the qubits on which the EJM is performed.}\label{FigConnTriangle}
\end{figure}
In Figure \ref{FigConnTriangle} the connectivity maps of the \textit{ibmq\_johannesburg} device and the \textit{ibmq\_montreal} device are displayed. The \textit{ibmq\_johannesburg} device was chosen for the triangle network experiment because of its ideal connectivity that allowed to perform the experiment directly on a ring of six qubits, while the \textit{ibmq\_montreal} device was chosen of its small readout and two-qubit gate errors. The particular colouring corresponds to the six qubits that were used on the respective devices. As there are no inputs in this setting, there is only one quantum circuit, which is shown in Figure \ref{FigCircTriangle}.
\begin{figure}[htb]
\centering
\includegraphics[width=0.8\columnwidth]{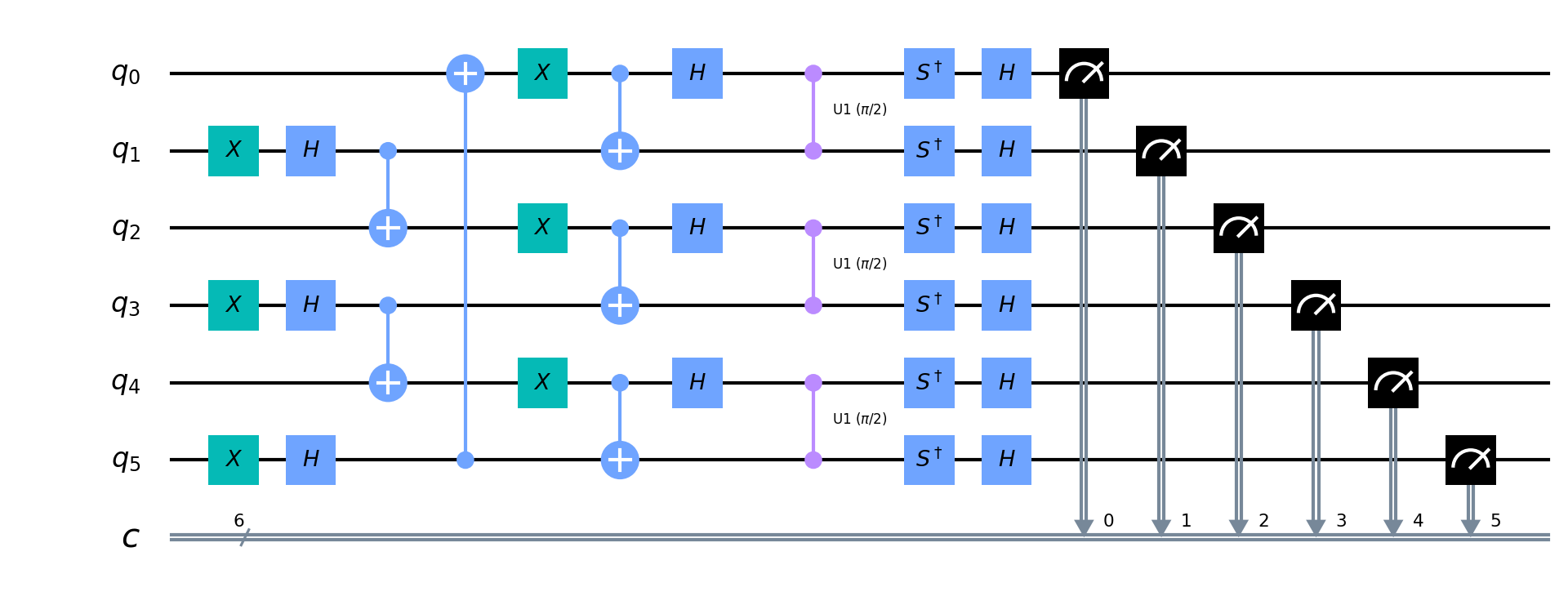}					
\caption{\textbf{Circuit implementing the quantum triangle-network experiment on six qubits.} In a first step, the maximally entangled singlet-state $\ket{\psi}=\frac{\ket{01}-\ket{10}}{\sqrt{2}}$ is created on each of the qubit pairs $(q_1,q_2)$, $(q_3,q_4)$, $(q_5,q_6)$, while in a second step, an EJM is performed on each of the qubit pairs $(q_0,q_1)$, $(q_2,q_3)$, $(q_4,q_5)$. See Figure \ref{FigConnTriangle} for the actually used qubits. }
\label{FigCircTriangle}
\end{figure}
}


\section{Statistical errors}\label{AppError}
We analyse the statistical errors in the star network experiment, the communication network experiment and the EJM bilocality experiment using a multinomial distribution. This is a direct generalisation of the binomial distribution: we consider $m$ independent trials, each of which result  in exactly one of $k$ different possible outcomes. The success probability of each outcome ($p_1,\ldots,p_k$) is given by the relative frequencies measured in the experiments. For such a distribution, the expected number of trials with outcome $o\in\{1,\ldots,k\}$ is given by $E_o=mp_o$ and the corresponding variance is $\sigma^2_{E_o}=mp_o(1-p_o)$. Thus, the standard deviation associated to a measured probability becomes
\begin{equation}
\sigma_{p_o}=\sqrt{\frac{p_o(1-p_o)}{m}}.
\end{equation}

\subsection{Communication network experiment}\label{AppErrorComm}
For the communication network, the final statistical error on $p_N^\text{win}$ admits a simple form, as we can directly propagate the independent event errors to the final winning probability $p_N^\text{win}\comm{=\frac{1}{4^{N}}\sum_{x,y \in \{0,1\}^N} p(b|x,y)}$,
\begin{align}
\sigma_{p_N^\text{win}}^2 &= \frac{1}{16^N} \sum_{x,y\ \in \{0,1\}^N} \sigma_o(b|x,y)^2\\
&= \frac{1}{16^Nm} \sum_{x,y\ \in \{0,1\}^N}  p(b|x,y)(1-p(b|x,y)).
\end{align}
Note that as here we are only interested in one ``winning element" for each setting, we only take one $b$ into account, so $b=b(x,y)$ is a function of $x$ and $y$.
\\
\\
Each circuit was implemented in 24576 shots ($N=2$), 8192 shots ($N=3,4$), 1024 shots ($N=5$), 128 shots ($N=6$), 32 shots ($N=7,8$), 16 shots ($N=9$) and 8 shots ($N=10$). \comm{Note, that while the number of shots might seem very small to determine the winning probability of measuring the correct one out of $2^{N}$ basis states, the fact that $p_N^\text{win}$ is the average over $4^{N}$ different setups allows us to still get a very good estimate not for each individual probability $p(b|x,y)$, but for the overall winning probability $p_N^\text{win}$.} This leads to a standard deviation (statistical error) no larger than $10^{-3}$ in the estimate of $p_N^\text{win}$  for $N<10$, as depicted in table \ref{TableFakeStd}.

\begin{table}[htb]
	\begin{tabular}{|c|cccccccc|}
		\hline
		N: \#qubits                                                                 & 2     & 3     & 4     & 5     & 6     & 7     & 8      & 9  \\ \hline
		\begin{tabular}[c]{@{}c@{}}Measured $p_N^\text{win}$   (\%)\end{tabular}         & 93.9 & 89.2 & 85.0 & 80.4 & 73.5 & 67.5 & 63.7 & 58.0  \\ \hline
		\begin{tabular}[c]{@{}c@{}} Standard deviation \end{tabular} & 3.8 e-4     & 4.3 e-4     & 2.5 e-4   & 3.9 e-4    & 6.0 e-4    & 6.3 e-4    & 3.2 e-4     & 2.3 e-4 \\ \hline
	\end{tabular}
	\caption{Standard deviations for the $N$-qubit communication network experiments on the \emph{ibmq$\_$montreal} quantum computer.}\label{TableFakeStd}
\end{table}

\subsection{Star network experiment}\label{AppErrorStar}
In the star network experiment, these errors on the individually measured probabilities were propagated to $\mathcal{S}_N$ via standard procedure. First, we need to consider the error propagated to the different $I_j$. Even though we assume independent errors on the statistics, the errors on $I_j$ can be correlated, which is why we need to consider the full variance-covariance matrix $\Sigma$,
\begin{equation}
\Sigma^{I} = \begin{pmatrix} 
\sigma_{11}^I & \sigma_{12}^I & \sigma_{13}^I & ...\\
\sigma_{12}^I & \sigma_{22}^I & \sigma_{23}^I & ...\\
\sigma_{13}^I & \sigma_{23}^I & \sigma_{33}^I & ... \\
... & ... & ... & ...
\end{pmatrix},
\end{equation}
with 
\begin{align}
\sigma_{ij}^I &= \frac{1}{4^N} \sum_{\bar{x},\bar{a},b} (-1)^{f_i(b)+f_j(b) +  g_i(\bar{x}) + g_j(\bar{x})} \sigma_o (\bar{a},b|\bar{x})^2,\\
&=\frac{1}{4^Nm} \sum_{\bar{x},\bar{a},b} (-1)^{f_i(b)+f_j(b) +  g_i(\bar{x}) + g_j(\bar{x})} p(\bar{a},b|\bar{x}) (1-p(\bar{a},b|\bar{x})),
\end{align}
which for the case $i=j$ simplifies to
\begin{align}
\sigma_{I}^2 := \sigma_{ii} &=\frac{1}{4^Nm} \sum_{\bar{x},\bar{a},b} p(\bar{a},b|\bar{x}) (1-p(\bar{a},b|\bar{x})).
\end{align}
In a second step we propagate the errors of the different $I_j$ to $\mathcal{S}_N$,
\begin{align}
\sigma^2_{\mathcal{S}_N} &= \frac{1}{N^2 4^{N-2}}\sum_{i,j=1}^{2^{N-1}} |I_i|^{1/N-1} |I_j|^{1/N-1} \sigma_{ij}^{I}\\
&=  \frac{1}{N^2 4^{N-2}} \sum_{j=1}^{2^{N-1}}  \left( |I_j|^{2/N-2}\sigma_I^2 +  2 \sum_{i<j}  |I_i|^{1/N-1} |I_j|^{1/N-1} \sigma_{ij}^{I} \right).
\end{align}
Note that the second term is negligible due to the alternating signs of $\sigma_{ij}^I$. \\
\\

For each circuit we have implemented approximately $1.2\times 10^5$ shots ($N=2,3$), $2\times 10^5$ shots ($N=4,5$) and $4.9\times 10^6$ shots ($N=6$). The standard deviations are depicted in table \ref{TableStarStd}.

\begin{table}[htb]
	\begin{tabular}{|c|ccccc|}
		\hline
		N: \#branches in network             & 2     & 3     & 4     & 5   &  6  \\ \hline
		\begin{tabular}[c]{@{}c@{}}Measured $\mathcal{S}_N$\end{tabular}         & 1.165 & 1.124 & 1.086 & 1.062 & 0.983  \\ \hline
		\begin{tabular}[c]{@{}c@{}}Standard deviation  \end{tabular}         & 1.7 e-3  & 1.1 e-3 & 6.1 e-4  & 5.1 e-4 & 3.7 e-4 \\ \hline
	\end{tabular}
	\caption{Standard deviations for $N$-branch star network nonlocality experiments on the \emph{ibmq$\_$almaden} quantum computer.}\label{TableStarStd}
\end{table}

The reason the standard deviation is decreasing with increasing $N$ in our experiments is that we increased the number of trials $m$.


\section{Measurement error mitigation}
While the present noise is a consequence of various kinds of errors, one of the most dominant factors is the noise due to readout errors. To account for those, we can post-process the data by applying measurement error mitigation \cite{Temme2017}. We note, however, that such a procedure is against the general spirit of quantum certification and quantum nonlocality experiments. By successively preparing and measuring all $2^N$ basis states, the transition probabilities of all basis states can be computed and captured in a calibration matrix. Given this calibration matrix a measurement filter can be created that can then be applied to the measurement statistics to mitigate the error. There are two different methods using Qiskit \cite{Qiskit} to create such a measurement filter: 
\begin{enumerate}[(i)]
	\item The \textit{``pseudo-inverse" method} which corresponds to the application of the pseudo-inverse of the calibration matrix. \label{pseudoi}
	\item The \textit{``least-squares" method}, where Sequential Least Squares Programming ( ``SLSQP") is used to find some mitigated data such that when the calibration matrix is applied it minimizes the difference to the raw data. \label{leastsq}
\end{enumerate}
While (\ref{pseudoi}) is faster to calculate, it can result in negative probabilities, which makes it less physical. (\ref{leastsq}) on the other hand will always lead to physical probabilities, but the optimization can take quite long for a large number of qubits. \\
\\
Note, that here the calibration matrices were not prepared on the same day as the actual experiments were performed. 
\comm{As the most significant contribution to measurement errors comes from thermal relaxation due to the coherence times of the qubits, there is a quite constant bias towards going from the excited to the ground state rather than the other way around. Thus, the calibration matrix stays very similar over time. Also, the correlated cross-talk errors do not change a lot over time, as they seem to depend strongly on the device connectivity and readout architecture.}
Nevertheless, the errors fluctuate in general non-negligible over time, and therefore a more accurate calibration matrix and possibly even better mitigated results could be achieved if it was prepared on the same day.

\subsection{Communication network experiment} \label{AppCommMitigation}

Due to computational limitations, we were only able to apply the measurement error mitigation method (\ref{leastsq}) for up to $N = 8$. The results are presented in Table \ref{TableFakeMit}. For the case $n=8$ we are able to certify 152 entangled operators using the mitigated data.

\begin{table}[htb]
	\begin{tabular}{|c|cccccccc|}
		\hline
		N: \#qubits                                                                 & 2     & 3     & 4     & 5     & 6     & 7     & 8      & 9  \\ \hline
		\begin{tabular}[c]{@{}c@{}}Measured $p_N^\text{win}$   (\%)\end{tabular}         & 93.9 & 89.2 & 85.0 & 80.4 & 73.5 & 67.5 & 63.7 & 58.0 \\ \hline
		\begin{tabular}[c]{@{}c@{}}\#certified entangled \\ basis elements\end{tabular} & 4     & 7     & 12    & 20    & 30    & 45    & 70     & 82 \\ \hline
		\begin{tabular}[c]{@{}c@{}}\ Meas. error mitigated  $p_N^\text{win}$ (least-squares) \end{tabular} & 98.7    & 98.0     &  95.8   & 93.3    & 88.3    & 83.8    & 79.6     & - \\ \hline
		\begin{tabular}[c]{@{}c@{}}\#certified entangled \\ basis elements\end{tabular} & 4     & 8     & 15    & 28    & 49    & 87    & 152     & -  \\ \hline
	\end{tabular}
	\caption{Mitigated results for $N$-qubit communication network experiments on the \emph{ibmq$\_$montreal} quantum computer.}\label{TableFakeMit}
\end{table}

\subsection{Star network experiment} \label{AppStarMitigation}
As we only managed to apply method (\ref{leastsq}) for up to $N = 5$ due to computational limitations, we have also applied method (\ref{pseudoi}) for all $N$. In Table \ref{TableStarMit} we present  the different results. While the pseudo-inverse seems to yield better results in general, we would like to stress again that the mitigated probability distribution can contain negative values and is therefore less physical, so the results have to be watched with care.
\begin{table}[htb]
	\begin{tabular}{|c|ccccc|}
		\hline
		N: \#branches in network             & 2     & 3     & 4     & 5   &  6  \\ \hline
		\begin{tabular}[c]{@{}c@{}}Measured $\mathcal{S}_N$\end{tabular}         & 1.165 & 1.124 & 1.086 & 1.062 & 0.984  \\ \hline
		\begin{tabular}[c]{@{}c@{}}Meas. error mitigated $\mathcal{S}_N$ (least-squares)\end{tabular}         & 1.408 & 1.339 & 1.343 & 1.297 & -  \\ \hline
		\begin{tabular}[c]{@{}c@{}}Meas. error mitigated $\mathcal{S}_N$ (pseudo-inverse)\end{tabular}         & 1.438 & 1.340 & 1.453 & 1.322 & 1.174  \\ \hline
	\end{tabular}
	\caption{Mitigated results for an $N$-branch star network nonlocality experiment on the \emph{ibmq$\_$almaden} quantum computer.}
	\label{TableStarMit}
\end{table}


\section{Details on the source-independent Bell inequalities of \cite{Tavakoli2014}}\label{AppStar}
Here we detail the Bell inequalities for the star network, originally derived in \cite{Tavakoli2014}, investigated in the main text. The inequalities read
\begin{equation}\label{ineqagain}
\mathcal{S}_N\equiv \frac{1}{2^{N-2}}\sum_{j=1}^{2^{N-1}} |I_j|^{1/N}\leq 1.
\end{equation} 
Here, $I_1,\ldots, I_{2^{N-1}}$ are linear combinations of correlators that involve all nodes in the network. These correlators are defined as 
\begin{equation}
\langle A_1^{x_1} \ldots A_N^{x_N}B_j\rangle=\sum_{\substack{a_1,\ldots,a_N\\ b_1,\ldots,b_N}}(-1)^{f_j(b)+\sum_{i=1}^N a_i}p(\bar{a},b|\bar{x}),
\end{equation}
where $f_j(b)$ maps the $N$-bit string $b$ into a single bit. Then, the $I_j$s are obtained from
\begin{equation}
I_j=\frac{1}{2^N}\sum_{\bar{x}} (-1)^{g_j(\bar{x})}\langle A_1^{x_1} \ldots A_N^{x_N}B_j\rangle,
\end{equation}
where $g_j(\bar{x})$ is a bit-valued function of the $N$-bit string $\bar{x}$. Thus, the inequality \eqref{ineqagain} requires us to specify the maps $\{f_j(b)\}_{j}$ and $\{g_j(\bar{x})\}_j$. We now list these functions for $N=2,3,4$, from which the extension the larger $N$ becomes clear.

For $N=2$, we have
\begin{align}
& f_1=b_1, \qquad  f_2=b_1\oplus b_2\oplus 1, \qquad g_1=0, \qquad g_2=x_1\oplus x_2.
\end{align}

For $N=3$, we have
\begin{eqnarray}
& f_1=b_1, \qquad f_2=b_1\oplus b_2\oplus 1, \qquad f_3=b_1\oplus b_3\oplus 1, \qquad f_4=b_1\oplus b_2\oplus b_3\oplus 1\\
& g_1=0, \qquad g_2=x_1\oplus x_2, \qquad g_3=x_1\oplus x_3, \qquad g_4=x_2\oplus x_3.
\end{eqnarray}
 
For $N=4$, we have
\begin{eqnarray}
& f_1=b_1, \qquad f_2=b_1\oplus b_2\oplus 1, \qquad f_3=b_1\oplus b_3\oplus 1, \qquad f_4=b_1\oplus b_4\oplus 1,\\
& f_5=b_1\oplus b_2\oplus b_3\oplus 1, \qquad f_6=b_1\oplus b_2\oplus b_4\oplus 1, \qquad f_5=b_1\oplus b_3\oplus b_4\oplus 1, \qquad f_5=b_1\oplus b_2\oplus b_3\oplus b_4\\
& g_1=0, \qquad g_2=x_1\oplus x_2, \qquad g_3=x_1\oplus x_3, \qquad g_4=x_1\oplus x_4,\\
& g_5=x_2\oplus x_3, \qquad g_6=x_2\oplus x_4, \qquad g_7=x_3\oplus x_4,  \qquad g_8=x_1\oplus x_2 \oplus x_3\oplus  x_4.
\end{eqnarray}

From the above, it is straightforward to extend the list also for larger values of $N$. See the original reference \cite{Tavakoli2014} for futher details.

\end{document}